\documentclass[aps, preprint, prl, superscriptaddress, showpacs,nobibnotes, longbibliography]{revtex4-1}
\usepackage{array,braket,mathtools}
\usepackage{graphicx}
\usepackage{bm, amsmath, siunitx}
\usepackage{dcolumn}
\usepackage[version=4]{mhchem}
\usepackage{natbib,xcolor}
\usepackage{comment}
\usepackage{xspace}

\newcommand{\degree}{$^\circ$}
\newcommand{\moire}{moir{\'e}\xspace}
\newcommand{\Moire}{Moir{\' e}\xspace}
\newcommand{\TaS}{TaS$_2$\xspace}
\newcommand{\SrTaS}{Sr$_6$TaS$_8$\xspace}
\newcommand{\SrTaSTaS}{(Sr$_6$TaS$_8$)$_{1+\delta}$(TaS$_2$)$_8$\xspace}
\newcommand{\STSnphi}{STS($n$,$\phi$)\xspace}
\newcommand{\q}{\textbf{\textit{q}}\xspace}
\newcommand{\qvector}{$q$-vector\xspace}
\newcommand{\aaxis}{\textit{a}-axis\xspace}
\newcommand{\baxis}{\textit{b}-axis\xspace}
\newcommand{\caxis}{\textit{c}-axis\xspace}

\newcommand{\DeltaMTau}{\ensuremath{\Delta M_\tau}\xspace}

\usepackage{ragged2e}
\usepackage{setspace}

\makeatletter
\long\def\@makecaption#1#2{%
  \vskip\abovecaptionskip
  \setstretch{0.9}
  \parbox{\linewidth}{%
    \small\justifying
    \noindent \textbf{Figure~\@arabic\c@figure\ \textbar}\ #2%
  }%
  \vskip\belowcaptionskip
}
\makeatother

\begin{document}
\title{Higher-dimensional Fermiology in bulk \moire metals}

\author{Kevin P. Nuckolls}
\thanks{These authors contributed equally to this work.}
\affiliation{Department of Physics, Massachusetts Institute of Technology, Cambridge, MA 02139, USA}
\author{Nisarga Paul}
\thanks{These authors contributed equally to this work.}
\affiliation{Department of Physics, Massachusetts Institute of Technology, Cambridge, MA 02139, USA}
\author{Alan Chen}
\affiliation{Department of Electrical Engineering and Computer Science, Massachusetts Institute of Technology, Cambridge, MA 02139, USA}
\author{Filippo Gaggioli}
\affiliation{Department of Physics, Massachusetts Institute of Technology, Cambridge, MA 02139, USA}
\author{Joshua P. Wakefield}
\affiliation{Department of Physics, Massachusetts Institute of Technology, Cambridge, MA 02139, USA}
\author{Avi Auslender}
\affiliation{Center for Nanoscale Systems, Harvard University, Cambridge, MA 02138, USA}
\affiliation{John A. Paulson School of Engineering and Applied Sciences, Harvard University, Cambridge, MA 02138, USA}
\author{Jules Gardener}
\affiliation{Center for Nanoscale Systems, Harvard University, Cambridge, MA 02138, USA}
\author{Austin J. Akey}
\affiliation{Center for Nanoscale Systems, Harvard University, Cambridge, MA 02138, USA}
\author{David Graf}
\affiliation{National High Magnetic Field Laboratory, Tallahassee, FL 32310, USA}
\author{Takehito Suzuki}
\affiliation{Department of Physics, Toho University, Funabashi, Japan}
\author{David C. Bell}
\affiliation{Center for Nanoscale Systems, Harvard University, Cambridge, MA 02138, USA}
\affiliation{John A. Paulson School of Engineering and Applied Sciences, Harvard University, Cambridge, MA 02138, USA}
\author{Liang Fu}
\affiliation{Department of Physics, Massachusetts Institute of Technology, Cambridge, MA 02139, USA}
\author{Joseph G. Checkelsky}
\thanks{checkelsky@mit.edu}
\affiliation{Department of Physics, Massachusetts Institute of Technology, Cambridge, MA 02139, USA}

\date{\today}
\maketitle

\textbf{
In the past decade, \moire materials have revolutionized how we engineer and control quantum phases of matter \cite{Balents2020, Mak2022, Nuckolls2024}. They are versatile platforms for strongly correlated electronic phenomena \cite{Cao2018a, Cao2018b, Cai2023, Lu2024} and support novel ferroelectric \cite{Yasuda2021, ViznerStern2021}, magnetic \cite{Song2021, Huang2023}, and superconducting states \cite{Zhao2023}. Among incommensurate materials \cite{Cummins1990}, \moire materials are aperiodic composite crystals \cite{VanSmaalen2007, Janssen2018} whose long-wavelength \moire superlattices enable tunable properties without chemically modifying their component layers. To date, nearly all reports of \moire materials have investigated van der Waals heterostructures assembled far from thermodynamic equilibrium ($T < 150$ \degree C) \cite{Balents2020, Mak2022, Nuckolls2024, Yankowitz2019}. Here we introduce a conceptually new approach to synthesizing high-mobility \moire materials in thermodynamic equilibrium. We report a new family of foliated superlattice materials \SrTaSTaS that are exfoliatable van der Waals crystals with atomically incommensurate lattices. Lattice mismatches between alternating layers generate \moire superlattices, analogous to those of 2D \moire heterobilayers, that are coherent throughout these crystals and are tunable through their synthesis conditions without altering their chemical composition. High-field quantum oscillation measurements map the complex Fermiology of these \moire metals \cite{Shoenberg1984, Abrikosov2017, Onsager1952, Lifshits1955, Alexandradinata2023, Leeb2025}, which can be tuned via the \moire superlattice structure. We find that the Fermi surface of the structurally simplest \moire metal is comprised of over 40 distinct cross-sectional areas, the most observed in any material to our knowledge. This can be naturally understood by postulating that bulk \moire materials can encode electronic properties of higher-dimensional superspace crystals in ways that parallel well-established crystallographic methods used for incommensurate lattices \cite{DeWolff1974, Janner1977, Janner1980a, Janner1980b}. More broadly, our work demonstrates a scalable synthesis approach potentially capable of producing large-area \moire materials for electronics applications and evidences a novel material design concept for accessing a broad range of physical phenomena proposed in higher dimensions \cite{Hazzard2023, Ozawa2019, Sugawa2018, Sakai2017, Cao2020}.
}

Crystalline order is typically the most stable configuration for atoms to form solids, but this is not always the case even within minerals found in nature \cite{Evans1968, Organova1973}. Incommensurate lattice crystals \cite{Cummins1990}, which lack translation symmetry while still supporting long-range order, are comprised of incommensurately modulated crystals (\textit{\textit{e.g.},} certain correlated oxides \cite{Dagotto2005, Comin2016, Yoshizawa1988, Birgeneau1989, Kivelson1998} and incommensurate ferroelectrics \cite{Yadav2016}), aperiodic composite crystals (\textit{e.g.}, misfit layer compounds \cite{Wiegers1996, Ng2022} and host-guest inclusion composites \cite{Harris1990}), and quasicrystals (\textit{e.g.}, icosahedral intermetallics \cite{Shechtman1984, Nelson1985, Goldman1993}). Atomically incommensurate materials (the latter two subclasses) evade the constraints of highly symmetric lattices found in periodic crystals. Furthermore, their electronic properties are difficult to capture using conventional band theory \cite{Azbel1979, Kohmoto1983, Kohmoto1986, Kollar1986, Zhang2015, SuarezMorell2010, Bistritzer2011, Wu2018, Wu2019} because they lack a well-defined Brillouin zone and a real-space unit cell under the crystal's physical dimensions. Instead, crystallographers describe incommensurate lattices using the mathematical framework of superspace crystals \cite{DeWolff1974, Janner1977, Janner1980a, Janner1980b}, which are emergent higher-dimensional periodic lattices derived from lower-dimensional incommensurate lattices that fully encode their superspace symmetries.

Recently, 2D \moire materials have emerged as a highly tunable class of incommensurate materials \cite{Balents2020, Mak2022, Nuckolls2024}. 2D \moire materials, like all aperiodic composites, are atomically incommensurate despite having well-defined local atomic coordinations. Like some aperiodic composites, the interference between lattice-mismatched crystalline layers produces a long-wavelength ``\moire superlattice'' modulation. Notably, \moire superlattices are tunable without altering the chemical composition of their layers (\textit{e.g.}, via twisting), and they support new low-energy properties not found in their constituent layers. This includes new band structures and Fermi surfaces \cite{SuarezMorell2010, Bistritzer2011}, new interlayer lattice or exchange couplings \cite{Yasuda2021, ViznerStern2021, Song2021, Huang2023}, and new correlated and topological phases \cite{Cao2018a, Cao2018b, Cai2023, Lu2024}.

Here we demonstrate a conceptually new approach to synthesizing high-mobility \moire materials using bulk synthesis techniques, contrasting existing methods that utilize the low-temperature assembly of mechanically exfoliated van der Waals flakes \cite{Balents2020, Nuckolls2024, Yankowitz2019}. We report a new family of aperiodic composite crystals, \SrTaSTaS for $\delta \approx 0.1$, that emulates the essential properties of lattice-mismatched heterobilayer \moire materials \cite{Mak2022}. Akin to a vertical stack of \SrTaS/\TaS \moire heterobilayers, these are also mechanically exfoliatable van der Waals crystals (see SI). Therefore, this family is a proof-of-concept for a potentially scalable approach for producing high-quality \moire materials for next-generation electronics using the tools of solid-state chemistry. The Fermiology of these bulk \moire metals, as mapped out through quantum oscillation measurements \cite{Shoenberg1984, Abrikosov2017, Lifshits1955, Alexandradinata2023, Leeb2025}, can be tuned via the crystal's tunable \moire superlattice. Furthermore, our measurements reveal a dense spectrum of quantum oscillation frequencies that are inconsistent with previously proposed non-Onsager mechanisms \cite{Onsager1952, Leeb2025}. Instead, this spectrum can be naturally understood by postulating that new oscillation frequencies stem from extremal Fermi-surface orbits that propagate into a synthetic superspace dimension generated by the \moire superlattice, a framework inspired by higher-dimensional superspace models of incommensurate crystallography \cite{DeWolff1974, Janner1977, Janner1980a, Janner1980b}. Our results pave the way towards using bulk \moire materials as superspace crystal platforms for accessing open theoretical proposals in higher dimensions \cite{Hazzard2023, Ozawa2019, Sugawa2018, Sakai2017, Cao2020}.

\begin{flushleft}\textbf{Bulk \Moire Materials from Aperiodic Composite Crystals}\end{flushleft}

\begin{figure}
	\includegraphics[width = \columnwidth]{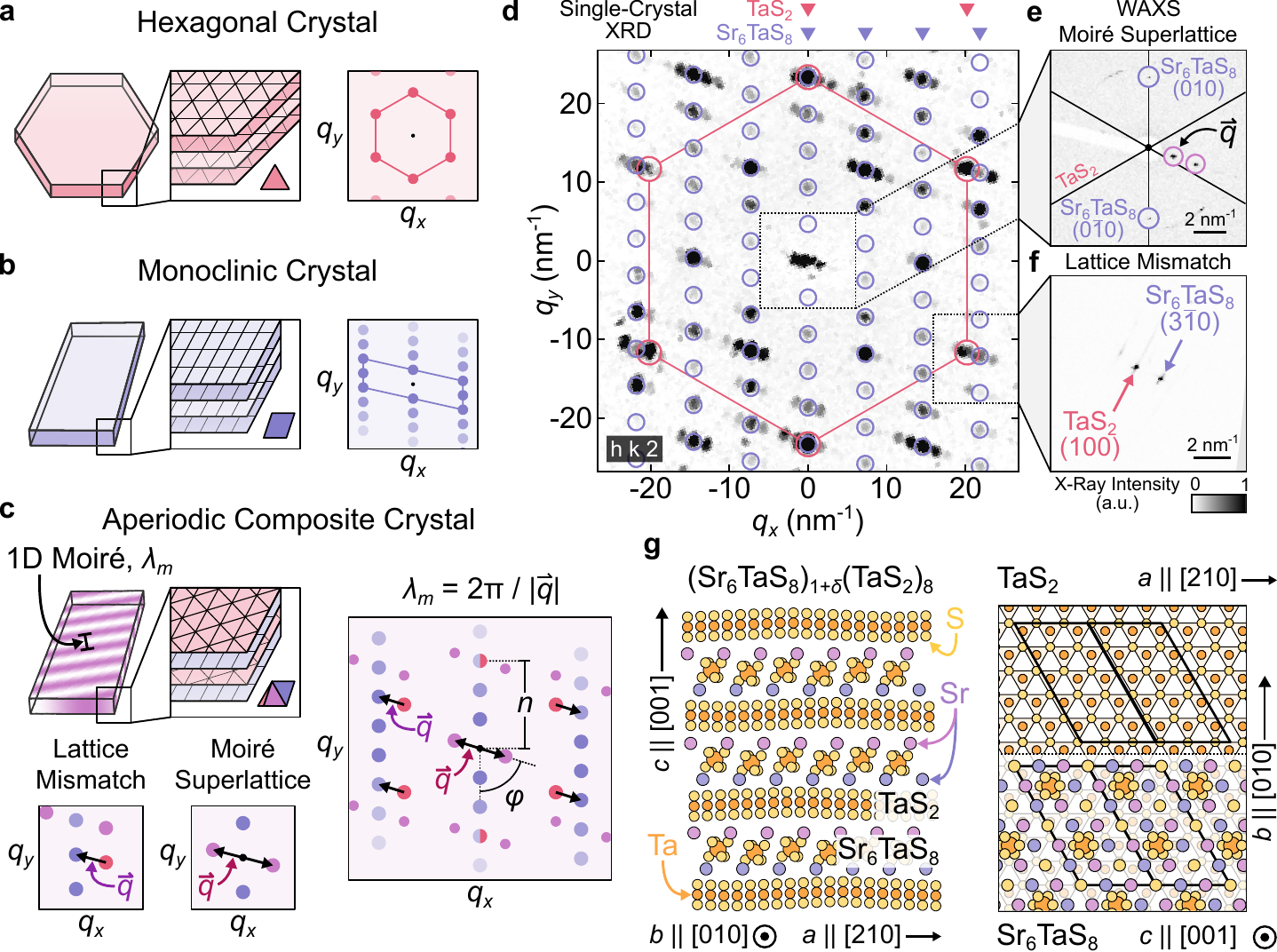}
    \caption{\label{Fig1} \textbf{Introduction to Bulk \Moire Materials.} \textbf{a,} Schematic of a hypothetical hexagonal crystal (left panel) and its expected diffraction pattern (right panel). \textbf{b,} Same as \textbf{a}, for a monoclinic crystal. \textbf{c,} Same as \textbf{a}, for a hypothetical aperiodic composite crystal (top left panel). X-ray and electron diffraction patterns (right panel) show a lattice-mismatch of Bragg vectors between neighboring layers (bottom left panel) and a long-wavelength \moire superlattice represented by a short \qvector (bottom center panel). The relevant features of this diffraction pattern can be captured by three quantities: the commensurate Bragg plane order $n$, and the \moire wavevector angle $\phi$, and the \moire wavevector length $|\boldsymbol{q}|$. \textbf{d,} Single-crystal diffraction (SCXRD) measurement of \SrTaSTaS, a new aperiodic composite compound. \textbf{e-f,} Zoomed-in wide-angle X-ray scattering (WAXS) showing the \moire superlattice and lattice-mismatch between \SrTaS and \TaS layers. \textbf{g,} Proposed highest symmetry model structure consistent with all structural and stoichiometric characterization measurements. Alternating \SrTaS and \TaS layers are atomically incommensurate with one another along the \aaxis, and they share commensurate Bragg planes along the \baxis. The black parallelogram outline denotes the intralayer unit cell of the \SrTaS layer, which is incommensurate with \TaS along the \aaxis. [hkl] crystallographic directions are defined with respect to the \TaS lattice.}
\end{figure}

Our material design principle is depicted in Figs. \ref{Fig1}a-c, showing two hypothetical layered compounds (left panels; Figs. \ref{Fig1}a,b) with hexagonal and monoclinic crystal systems and their expected in-plane diffraction patterns (right panels; Figs. \ref{Fig1}a,b). Foliated superlattice materials \cite{Devarakonda2020, Ma2022, Devarakonda2024, Lin2025} may result from the natural intergrowth of these dissimilar crystals, although not by respecting the competing symmetry constraints of both crystal systems simultaneously. However, an accommodative reaction product that mitigates interlayer lattice frustration could satisfy a partial subset of these symmetry constraints. For example, the schematically depicted rightmost Bragg planes of the hexagonal layers (center panel; Fig. \ref{Fig1}a) and monoclinic layers (center panel; Fig. \ref{Fig1}b) could align to form a new foliated superlattice material (Fig. \ref{Fig1}c). Here, hexagonal and monoclinic lattices are commensurate along the $q_y$ in-plane lattice direction (\baxis) and share a Bragg reflection peak (right panel; Fig. \ref{Fig1}c). However, this necessarily enforces the two interleaved layer types to be atomically incommensurate along the orthogonal $q_x$-direction / \aaxis in the absence of enormous lattice relaxation or reconstruction effects. Hence, the result is an aperiodic composite crystal \cite{Cummins1990, VanSmaalen2007, Janssen2018}, accessible using a salt-catalyzed reaction known to favor foliated superlattice lattice motifs over competing non-superlattice polymorphs \cite{Devarakonda2020, Ma2022, Devarakonda2024, Lin2025}.

Importantly, X-ray diffraction techniques like single-crystal diffraction (SCXRD) and wide-angle X-ray scattering (WAXS) can identify two distinguishing signatures of aperiodic composite lattices. For $|\boldsymbol{k}| \gg 0$ (bottom left panel; Fig. \ref{Fig1}c), lattice-mismatch incommensurability between adjacent layers appears as a superposition of the diffraction patterns from each layer (right panel; Fig. \ref{Fig1}c), where Bragg vectors $\boldsymbol{G}_1$ and $\boldsymbol{G}_2$ from different layer types (red and purple spots) are separated by a small wavevector $\boldsymbol{q} = \boldsymbol{G}_1 - \boldsymbol{G}_2$. Near $\boldsymbol{k} = 0$ (bottom center panel; Fig. \ref{Fig1}c), \q itself reveals the modulation of a \moire superlattice derived from the system's lattice-mismatch (distinct from the out-of-plane foliated superlattice). Both signatures are typical of 2D \moire materials \cite{Yoo2019, Kazmierczak2021}. 

SCXRD measurements (Fig. \ref{Fig1}d) of \SrTaSTaS, a new foliated superlattice material, show Bragg reflections confined to specific values of $q_x$ (arrow marker; Fig. \ref{Fig1}d). Most peaks can be attributed to the hexagonal transition metal dichalcogenide (TMD) monolayers (\TaS, red circles) or the monoclinic ``spacer'' monolayers (\SrTaS, purple circles). Superimposed peaks from different lattices (concentric circles; $q_y \approx 22$ nm$^{-1}$) indicate that the TMD and spacer layers share commensurate Bragg planes along the \baxis. However, many diffraction peaks cannot be attributed to either lattice individually. Higher-resolution WAXS data show the Bragg reflections of different layers separated by a small momentum \q (Fig. \ref{Fig1}f), while this same \q and its higher harmonics appear near the origin (Fig. \ref{Fig1}e) and as satellites near dominant Bragg reflections (Fig. \ref{Fig1}d; see SI). These features are indicative of a 1D \moire superlattice with a \moire wavelength of $\lambda_m = 2 \pi / |\boldsymbol{q}| \approx 4.13$ nm, comparable to $\lambda_m$ observed in 2D \moire heterobilayers \cite{Mak2022}. 

The sharpness of diffraction peaks associated with \q and its influence on scattering extinction conditions (see SI) demonstrate that naturally grown \moire superlattices show minimal structural disorder throughout the bulk of millimeter-sized crystals. Moreover, nearly identical diffraction patterns obtained from any crystal within a batch (see SI) evidences the reproducibility and potential scalability of our approach. In contrast, nominally identical device assembly procedures produce 2D \moire superlattices with twist-angle variations and strain gradient profiles \cite{Lau2022}, highlighting the advantages of our synthesis approach in addressing reproducibility. Given the symmetries of these diffraction patterns (Fig. \ref{Fig1}d) and those of related measurements (see SI), we construct the highest symmetry model structure consistent with all characterization observables (Fig. \ref{Fig1}g). \SrTaSTaS manifests our material design principle (Figs. \ref{Fig1}a-c), consisting of the natural intergrowth of lattice-mismatched monolayers in a moir{\' e}-modulated aperiodic composite structure.

\begin{flushleft}\textbf{Tuning \Moire Superlattices via Synthesis Parameters}\end{flushleft}

\begin{figure}
	\includegraphics[width = \columnwidth]{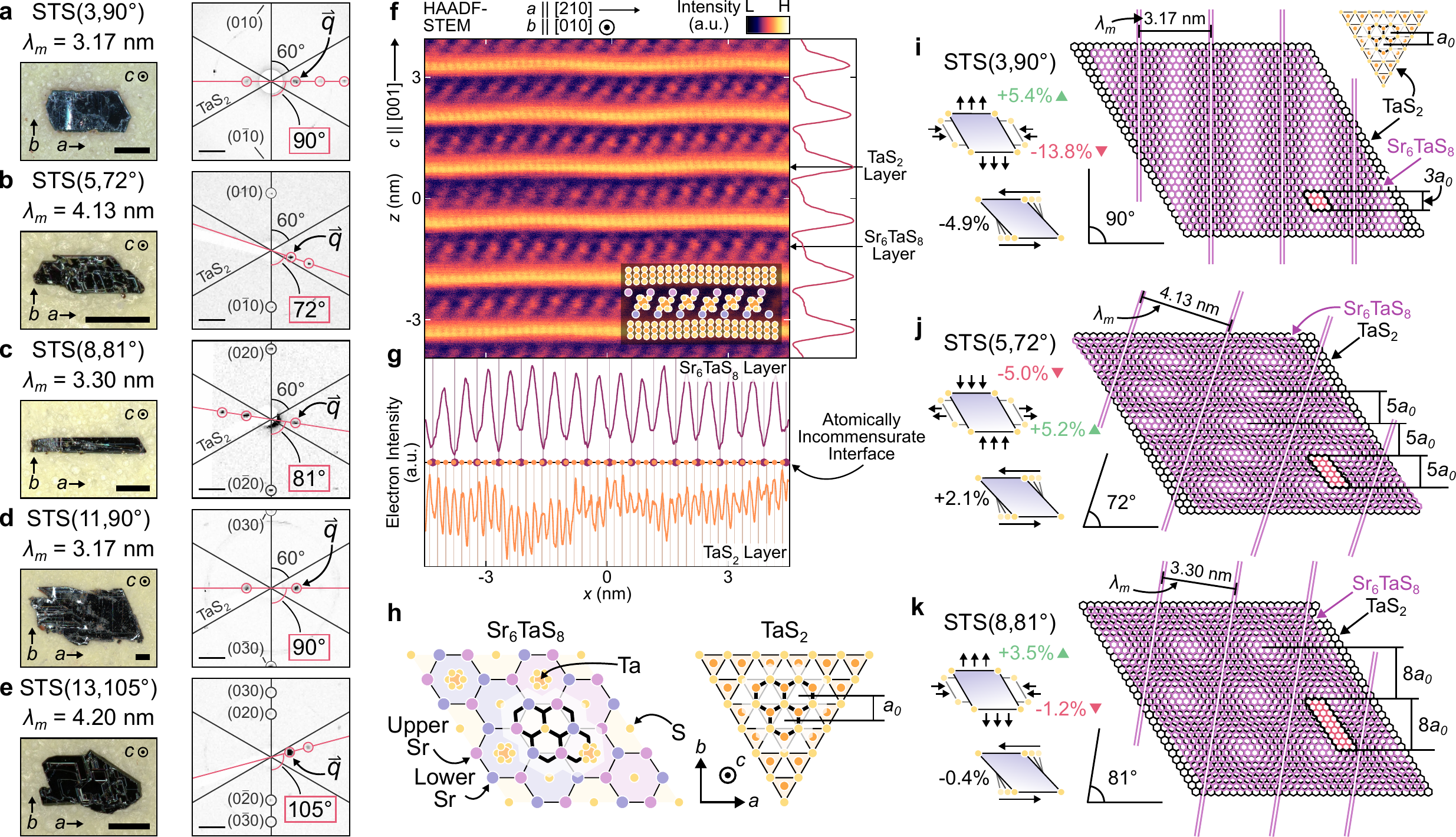}
    	\caption{\label{Fig2} \textbf{Tunable 1D \Moire Superlattices and 2D \Moire Approximants.} \textbf{a-e,} Optical images (left panels) and wide-angle X-ray scattering measurements (right panels) of five related families of bulk \moire materials. The spacer \SrTaS layer in each family ``snaps'' into $q_y$-oriented commensurate Bragg planes that 3- (\textbf{a}), 5- (\textbf{b}), 8- (\textbf{c}), 11- (\textbf{d}), and 13-tuple (\textbf{e}) the \TaS unit cell in this direction. Scale bars are 0.2 mm and 2 nm$^{-1}$. \textbf{f,} High-angle annular dark-field scanning transmission electron microscopy (HAADF-STEM) cross-sectional image of STS(8,81\degree) (\textbf{c}), which directly visualizes how neighboring \TaS and \SrTaS monolayers are mutually incommensurate with each other and are \moire-modulated in the \textit{z}-direction. \textbf{g,} Line cuts of the HAADF-STEM image in \textbf{f}. Neighboring \SrTaS and \TaS layers are atomically incommensurate with one another. \textbf{h,} Simplified visualizations of \SrTaS and \TaS lattices. \textbf{i-k,} Large field-of-view, simplified visualizations of representative \SrTaS / \TaS \moire patterns, realized via tunable interlayer couplings between \SrTaS and \TaS layers. Commensurate Bragg plane conditions ($a_0$-multiple right labels), the \moire wavelengths $\lambda_m$ (top labels), and their rotational alignment $\phi$ with respect to the \TaS lattice (bottom left labels) are all extracted quantitatively from X-ray diffraction measurements to construct these visualizations. Spacer unit cells are shown by black-outlined red-filled parallelograms. All percentages represent measured deviations of the lattice compression (red numbers), expansion (green numbers), and row-sliding shear (black numbers) of the \SrTaS layer lattice relative to an average spacer lattice model. STS(3,90\degree) represents the simplest structural member of the family with a purely 1D \moire superlattice, while STS(5,72\degree) and STS(8,81\degree) are representative of structures that are strictly 1D \moire superlattices, but are akin to 2D \moire approximants due to additional lattice aliasing in the commensurate \baxis.}
\end{figure}

An essential feature of \moire materials is their tunability without chemically altering their layers, typically achieved by rotationally misaligning layers manually during the device assembly process \cite{Balents2020, Mak2022, Nuckolls2024, Yankowitz2019}. We demonstrate that synthesis parameters can offer comparable tunability in bulk \moire materials. Figures \ref{Fig2}a-e show representative crystals from five distinct groups of bulk \moire materials, all ternaries of Sr-Ta-S accessed via temperature sequence and precursor stoichiometries variations (see Methods). Figure \ref{Fig2}b represents the class discussed in Figs. \ref{Fig1}d-g. Top-down images show the quasi-1D morphology of these ternary crystals (Figs. \ref{Fig2}a-e) with crystal facets that optically differentiate their longer \aaxis (horizontal) from their shorter \baxis (vertical) (see SI). X-ray diffraction reveals five parent structures, where the spacer layers ``snap'' into $q_y$-oriented (\baxis) commensurate Bragg planes that $3$- (a), $5$- (b), $8$- (c), $11$- (d), and $13$-tuple (e) the \TaS unit cell in this direction. In the orthogonal incommensurate direction (\aaxis), a wide spectrum of 1D \moire superlattices are realized with various \moire wavelengths $\lambda_m$ ($3$ nm to $4.2$ nm) and orientation angles $\phi$ ($\pm 20$\degree) near the \aaxis. Additionally, different synthesis conditions produce different \moire superlattices that even share the same commensurate Bragg condition (see SI). Each structure is denoted \STSnphi, where $n$ labels the commensurate Bragg plane order and $\phi$ defines the \moire \qvector angle relative to the compound's \baxis.

A natural hypothesis is that crystals grown under different synthesis conditions form chemically distinct ternaries in a complex phase space. However, stoichiometric probes like wavelength-dispersive X-ray spectroscopy (WDS) rule this out. Across $23$ crystal batches, WDS measurements show that the Sr:Ta molar ratios in these compounds are very similar, deviating from the mean by a maximum of $5.1\%$ (most within 3\%), comparable to the tool's experimental error (sulfur stoichiometries were inconclusive; see SI). This suggests that all of these foliated superlattice materials are the intergrowth of the same two monolayer crystals, \TaS and \SrTaS, with different \moire superlattice structures. To better understand this tunable \moire structure, we conducted high-angle annular dark-field scanning transmission electron microscopy (HAADF-STEM; Fig. \ref{Fig2}f) on cross-sectional lamellae of STS(8,81\degree), where the diffraction symmetries are more complex than those of STS(5,72\degree) (Fig. \ref{Fig2}b). HAADF-STEM images closely match features of the model structure previously proposed (Fig. \ref{Fig1}g) and additionally visualize the crystal's out-of-plane \moire-modulation along the length of these dendritic crystals. Hence, HAADF-STEM also suggests that chemically identical TMD and spacer layers intergrow to form a tunable variety of in-plane \moire superlattices.

Given these observations, we propose a comprehensive structural solution that builds bulk \moire crystals modularly from a related commensurate lattice motif (see SI) \cite{Khasanova2002}. This solution is consistent with all characterization data and can naturally explain the five structural families in Figs. \ref{Fig2}a-e and many other subtle observations (see SI for discussion). Subtle variations in interlayer heterostrain and heteroshear independently change $n$ and $\phi$, stabilizing different \STSnphi compounds (see SI). Furthermore, interlayer heteroshear in bulk \moire materials can be thought of as a continuously tunable analogue to twist angle in 2D \moire materials (see SI).

Further insights stem from abstracting the \TaS and \SrTaS lattices as simplified hexagonal and monoclinic lattices (Fig. \ref{Fig2}h), respectively, which are strained and sheared according to lattice parameters quantitatively extracted from WAXS (Figs. \ref{Fig2}i-k; see SI for details). These structures are strictly only atomically incommensurate along their \aaxis (horizontal); however, Figs. \ref{Fig2}j,k mimic strained 2D \moire superlattices in ways that parallel how so-called ``crystalline approximants'' mimic quasicrystalline lattices \cite{Goldman1993}. Whether the electronic properties of these materials are governed by an approximate 2D \moire superlattice or a strictly 1D \moire superlattice remains unresolved at this time and will be investigated in future work. 

\begin{flushleft}\textbf{Quantum Oscillation Maps of Tunable Fermiology}\end{flushleft}

The electronic properties of \moire materials can be controlled by the structure of their \moire superlattices \cite{Balents2020, Mak2022, Nuckolls2024}, a mechanism we probe directly through quantum oscillation maps of the Fermiology \cite{Shoenberg1984, Abrikosov2017, Lifshits1955} of bulk \moire metals for fields $\mu_0 H$ rotated by a polar angle $\theta$ within the \textit{bc}-plane. High-field torque magnetization and magnetotransport measurements (see SI) reveal a dense cascade of low-frequency oscillations onsetting in these \moire compounds near $\mu_0 H = 2 \text{ T}$ (see SI). We first attempt to isolate the \moire superlattice's influence upon Fermiology by comparing quantum oscillations in ``\moire cognate'' compounds, STS(8,81\degree) and STS(8,90\degree). These compounds are nearly structurally identical (see SI), but importantly their \moire $q$-vectors point in slightly different directions ($\phi = 81$\degree or $90$\degree) due to different degrees of interlayer heteroshear established under different growth conditions (see Methods).

\begin{figure}
	\includegraphics[width = \columnwidth]{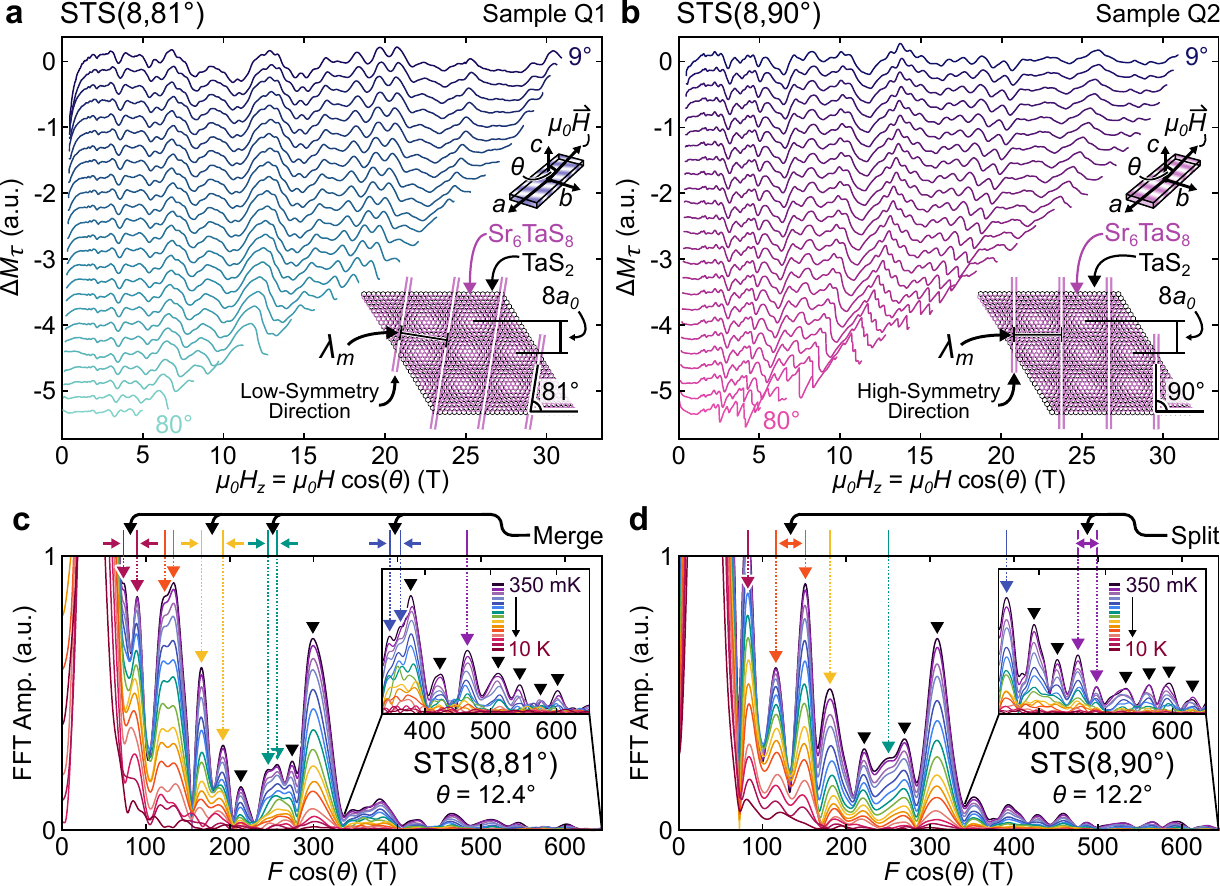}
	\caption{\label{Fig3} \textbf{Tunable Fermiology in \Moire Cognate Crystals.} \textbf{a,} High-field torque magnetization measurements of quantum oscillations in STS(8,81\degree) for the field $\mu_0 H$ rotated by a polar angle $\theta$ within the crystal's \textit{bc}-plane and scaled by the out-of-plane component $\mu_0 H \cos{\theta}$. The \moire \qvector in STS(8,81\degree) points along a non-crystallographic, low-symmetry direction, 81\degree from its \baxis. \textbf{b,} Same as \textbf{a}, for STS(8,90\degree), which is related to STS(8,81\degree) via different degrees of interlayer hetero-shear. The \moire \qvector in STS(8,90\degree) points along a crystallographic, high-symmetry direction, 90\degree from its \baxis (\textit{i.e.}, along the \aaxis). \textbf{c-d,} FFTs of the temperature-dependent background-subtracted torque magnetization \DeltaMTau of STS(8,81\degree) (\textbf{c}) and STS(8,90\degree) (\textbf{d}), showing cascades of low-frequency oscillations. Careful comparisons of these frequencies suggest that nearby Fermi pockets merge with one another (\textbf{c}) and split away from one another (\textbf{d}) as the layers of these bulk \moire materials shear against one another to sweep the \moire superlattice from a low-symmetry direction (81\degree) to a high-symmetry direction (90\degree) with respect to the \aaxis. Insets: Zoomed-in plots to show additional peaks in the FFT with lower FFT amplitudes.}
\end{figure}

At high magnetic fields ($\mu_0 H > 15 \text{ T}$), torque magnetization measurements (\DeltaMTau; Figs. \ref{Fig3}a,b) reveal de Haas-van Alphen oscillations that pull inwards to lower fields $\mu_0 H$ cos($\theta$) with increasing $\theta$, signifying corrugated quasi-2D Fermi pockets that narrow near the Brillouin zone boundary (see SI for more data). FFTs of the temperature-dependent torque magnetization (Figs. \ref{Fig3}c,d) of these \moire cognates exhibit similar amplitude profiles and frequency distributions, but they are clearly distinguishable. Importantly, by shearing the \moire superlattice from low- to high-symmetry directions, many neighboring FFT peaks appear to merge into sharp, isolated ones (top labels; Fig. \ref{Fig3}c), while a few broader peaks split into well-separated ones (top labels; Fig. \ref{Fig3}d) (see SI for direct comparison). Through Onsager's relation \cite{Onsager1952}, this suggests that the \moire superlattice direction reshapes and resizes Fermi pockets in these bulk crystals. Some of this data can be plausibly explained further by simple Brillouin-zone folding arguments (see SI), but this interpretation neglects several key observations that invite a more careful theoretical treatment.

\begin{figure}
	\includegraphics[width = 0.9\columnwidth]{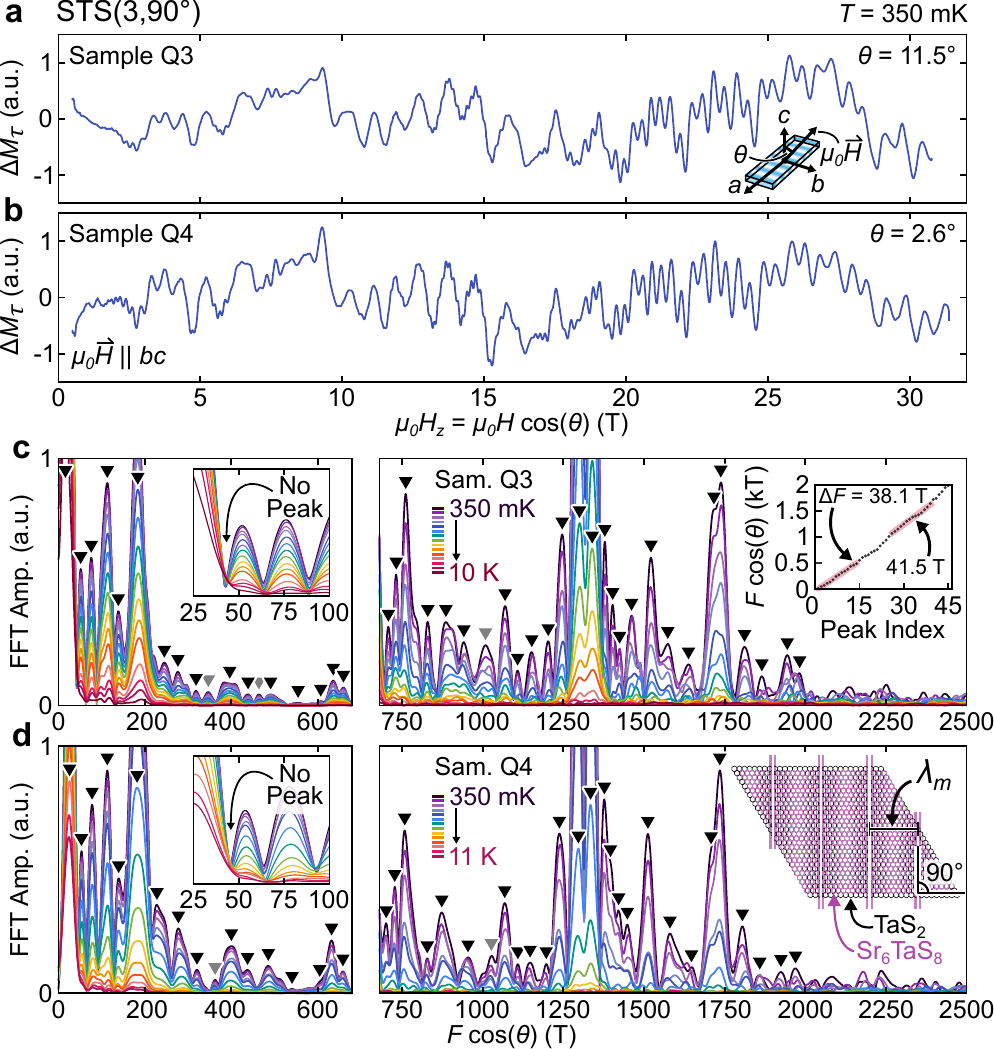}
	\caption{\label{Fig4} \textbf{de Haas-van Alphen Frequency Comb in the Fermiology of a Simple \Moire Metal.} \textbf{a-b,} High-field torque magnetization measurements of quantum oscillations in two crystals of STS(3,90\degree) for the field $\mu_0 H$ rotated by a polar angle $\theta$ within the crystal's \textit{bc}-plane and scaled by the out-of-plane component $\mu_0 H \cos{\theta}$. STS(3,90\degree) is a simple, purely 1D \moire metal with its \moire \qvector pointed along its incommensurate \aaxis (right inset of \textbf{d}). \textbf{c-d,} FFTs of the temperature-dependent background-subtracted torque magnetization \DeltaMTau of each sample of STS(3,90\degree), showing an abundance of oscillations that are all roughly linearly spaced in frequency. FFT amp. scaled in right panels by 12.2$\times$ (\textbf{c}) and 8$\times$ (\textbf{d}) for clarity. Black triangle markers identify FFT peaks that are closely shared between data sets. Grey triangle markers identify FFT peaks that are either at slightly different frequencies between data sets or are only observed in only one data set. Left insets: Zoomed-in plots of the ultra-low frequency regime of the FFT. Right inset of \textbf{c}: Extracted FFT peak frequencies as a function of peak index, showing a series of linearly spaced frequencies spaced by roughly $\Delta F \cos{\theta} \approx 40 \text{ T}$, a value not represented by any observed peak in the low-frequency limit. Fitted peak index regions (red shaded regions) identify subsequences of FFT peaks that appear as clear combs of FFT peaks with relatively similar amplitudes. A linear fit of the entire domain similarly yields linear peak spacings of $\Delta F \cos{\theta} \approx 44.6 \text{ T}$ with $R^2 = 0.996$, a value again not represented by any observed peak in the low-frequency limit.}
\end{figure}

These discrepancies are particularly apparent in the complex Fermiology of STS(3,90\degree), the structurally simplest \moire metal studied here. In STS(3,90\degree), \SrTaS and \TaS lattices lie in atomic register along the crystal's \baxis (Fig. \ref{Fig2}i), tripling the \TaS unit cell along this direction, and a 1D \moire superlattice with $\lambda_m = 3.17$ nm points along the crystal's incommensurate \aaxis. This separates the momentum scales of commensurate and incommensurate Fermi surface reorganization effects and enables us to isolate the salient consequences of a purely 1D \moire superlattice. Figures \ref{Fig4}a,b show the background-subtracted torque magnetization \DeltaMTau in two representative crystals up to $\mu_0 H = 31.5 \text{ T}$ . Notably, STS(3,90\degree) has moderately high-frequency oscillations absent in related compounds (see SI for discussion). FFTs of the temperature-dependent torque magnetization (Figs. \ref{Fig4}c,d) uncover a rich spectrum of de Haas-van Alphen frequencies. STS(3,90\degree) also hosts a cascade of low-frequency oscillations, but distinctive combs of FFT peaks extend up to approximately $2{,}000 \text{ T}$. Over $40$ FFT peaks appear at nearly identical frequencies in both samples (black triangle markers), the most of any metal to our knowledge (see SI). These observations are robust under various signal processing windows, especially those that suppress spectral leakage and higher harmonics, and to field windowing that includes the high-field regime (see SI for full analysis). Using a local-fit peak finding algorithm (see SI), we extract the frequencies of FFT peaks, which vary roughly linearly with peak index with an approximate spacing of $\Delta F \cos{\theta} \approx 40 \text{ T}$ (right inset; Fig. \ref{Fig4}c), a frequency not represented in the ultra-low frequency regime (left insets; Figs. \ref{Fig4}c,d).

The Fermiology of this simple \moire metal has several seemingly puzzling qualities. First, it seems counterintuitive to find sharp oscillation frequencies in an incommensurate metal without a well-defined Brillouin zone \cite{Zhang2015}. Furthermore, their peak widths are comparable to those of crystalline metals. Second, the approximate linear frequency spacing of these peaks exhibits a markedly higher degree of order than what one might expect of the Fermi surface of any multiband metal. Finally, the extremal areas of these oscillations sum to a much larger area ($\approx 41.7 \text{ kT}$) than the approximate Brillouin zone size for this material ($\approx 1.5 \text{ kT}$; see SI). There are several non-Onsager mechanisms \cite{Onsager1952, Leeb2025} capable of generating new oscillation frequencies unassociated with an extremal Fermi-surface orbit (\textit{e.g.}, magnetic breakdown \cite{Everson1987, Hill1997}, quasiparticle lifetime oscillations \cite{Huber2023}, Weiss oscillations \cite{Beenakker1989, Gerhardts1989, Devarakonda2024}), but all are unlikely explanations given the absence of a $40 \text{ T}$ peak in the FFT (insets; Figs. \ref{Fig4}c,d), the numerous generations of FFT peaks with similar amplitudes, and their temperature dependence (see SI for full discussion).

\begin{flushleft}\textbf{\Moire Metals with Synthetic Superspace Dimensions}\end{flushleft}

\begin{figure}
	\includegraphics[width = 0.87\columnwidth]{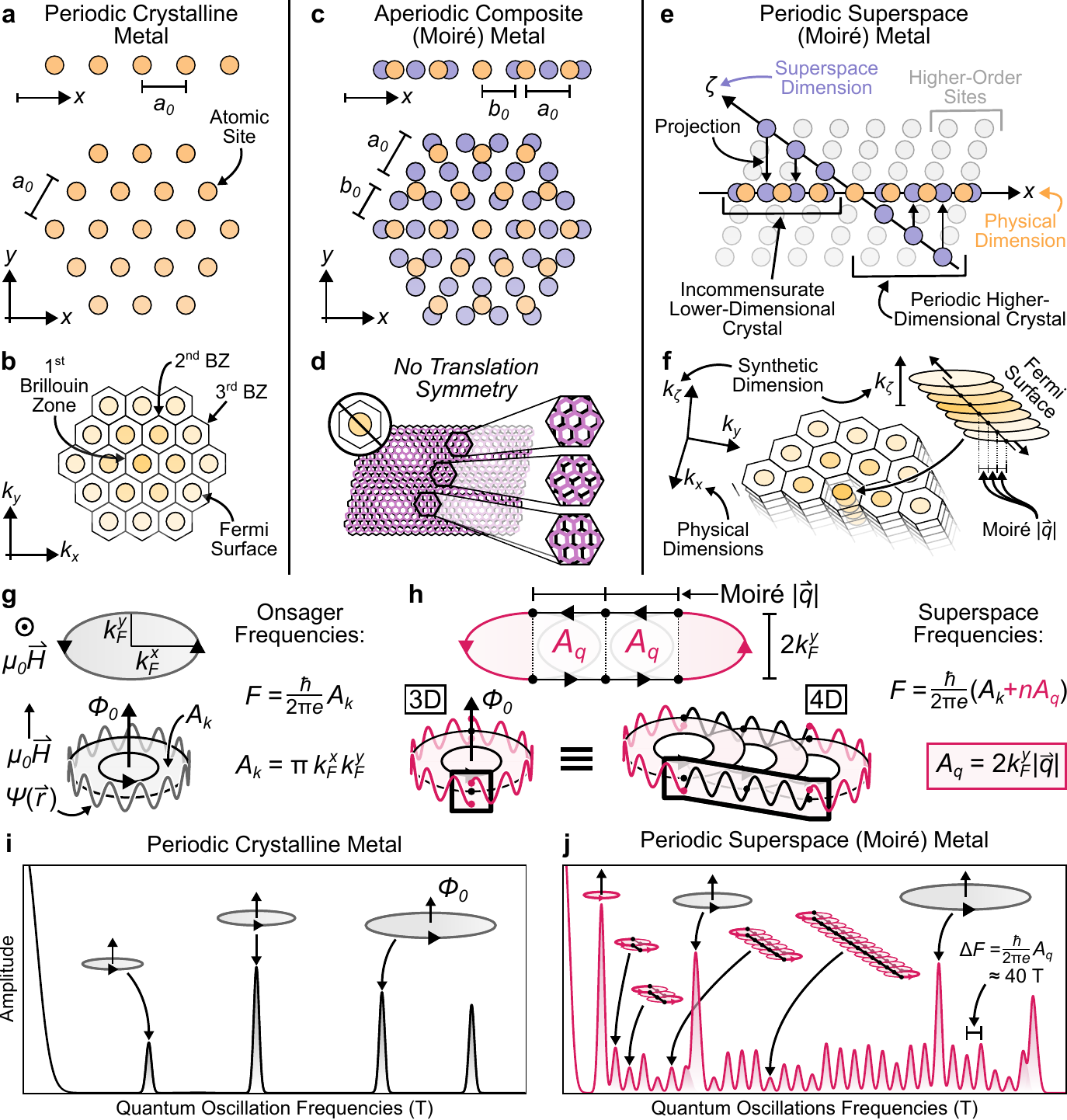}
	\caption{\label{Fig5} \textbf{\Moire Metals as Quantum Materials with Synthetic Superspace Dimensions.} \textbf{a-b,} Schematic of the periodic lattice (\textbf{a}) and the quasi-2D Fermi surface (\textbf{b}) of a quasi-2D crystalline metal. \textbf{c,} Schematic a quasi-2D aperiodic composite / \moire metal lattice. \textbf{d,} Schematic of the violated translation symmetry of the aperiodic composite / \moire lattice, which prohibits a complete description of this metal’s Fermiology in 2D. \textbf{e,} Schematic of the emergent, higher-dimensional, periodic superspace lattice described by a lower-dimensional aperiodic composite lattice. By elevating the quasi-2D aperiodic lattice into a new superspace dimension $\zeta$, an emergent translation symmetry is restored in ($2 + 1$)D. \textbf{f,} Schematic of the resulting ($2 + 1$)D Fermi surface, leveraging the restored translation symmetry. Stacks of quasi-2D Fermi surfaces are linked along a new discrete synthetic dimension $k_\zeta$. \textbf{g,} Schematic of an extremal Fermi surface orbit in a crystalline metal, whose associated frequency is given by Onsager's relation. The wavy line depicts the phase of the electronic wavefunction, which is $2 \pi$-periodic. \textbf{h,} Same as \textbf{g}, for a \moire metal. Superspace cyclotron orbits have phase discontinuities in 3D, but are continuous and $2 \pi$-periodic in 4D. A linear sequence of frequencies stem from superspace Fermi orbits that coherently hop into the synthetic dimension. \textbf{i-j,} Schematics of the oscillations frequencies and cyclotron orbits in a quasi-2D crystalline metal (\textbf{g}) and a ($2 + 1$)D superspace \moire metal (\textbf{h}). Cyclotron orbits in the ($2 + 1$)D superspace \moire metal propagate by discrete momentum quanta in the synthetic dimension, producing a dense sequence of roughly linearly spaced de Haas-van Alphen oscillations frequencies.}
\end{figure}

These initially perplexing observations can be naturally understood by drawing inspiration from a closely related solved problem concerning the crystallography of incommensurate lattices. Prior to $1960$, all solid matter was believed to be either crystalline (Figs. \ref{Fig5}a,b) or amorphous \cite{Janssen1988}. This viewpoint was challenged by discoveries of aperiodic composites \cite{Brown1974, Johnson1976, Pouget1978} (Fig. \ref{Fig5}c) and quasicrystals \cite{Shechtman1984, Nelson1985}, which exhibit long-range order without periodic unit cells \cite{Cummins1990}. Without translation symmetry, these lattices cannot support Bloch electronic bands in their physical dimensions (Fig. \ref{Fig5}d), although effective theories often suffice \cite{Azbel1979, Kohmoto1983, Kohmoto1986, Kollar1986, Zhang2015, SuarezMorell2010, Bistritzer2011, Wu2018, Wu2019}. Counterintuitively, these crystals exhibit dense arrays of sharp diffraction peaks despite not having a well-defined unit cell \cite{VanSmaalen2007, Janssen2018} with peaks widths comparable to those of periodic crystals \cite{Goldman1993}. This motivated the development of higher-dimensional superspace crystal models \cite{DeWolff1974, Janner1977} as a unifying framework for describing incommensurate lattices. In these models, long-range order in incommensurate lattices stems from their description as projections of a higher-dimensional periodic superspace lattice (Fig. \ref{Fig5}e) \cite{Janner1980a, Janner1980b, VanSmaalen1991}. Via this construction, the emergent symmetries of superspace crystals are fully encoded in lower-dimensional incommensurate lattice representations. By comparing our present problem, which aims to explain a dense array of sharp de Haas-van Alphen frequencies in a \moire metal, with this related problem from incommensurate crystallography, we propose that both are similarly rectified by elevating incommensurate structures into emergent superspace dimensions. Such approaches have previously explained static properties of solids, such as atomic lattices and charge densities \cite{VanSmaalen2007, Janssen2018}, but here we introduce how they apply to dynamic electronic phenomena, like the Fermi surfaces and cyclotron orbits of itinerant charge carriers.

We propose a momentum-space network model derived from a Lifshitz-Kosevich treatment of quasi-2D metals with incommensurate 1D \moire modulations (see SI for details). Its results are intuitively understood as describing quasi-2D Fermi pockets linked at momenta defined by the \moire \qvector along an emergent discrete dimension $k_\zeta$ (Fig. \ref{Fig5}f). This mirrors synthetic dimensions encoded by internal degrees of freedom in recent cold-atom experiments (\textit{e.g.}, via hyperfine states) \cite{Ozawa2019, Boada2012, Celi2014, Mancini2015, Lohse2018, Zilberberg2018, Lustig2019,Dutt2020}. In understanding aperiodic composite / \moire lattices in an emergent higher-dimensional space, we regain translation symmetry and therefore regain the ability to describe their properties in terms of Bloch bands of electronic states \cite{Lu1987}.

In this picture, a Fermi pocket of the unmodulated structure with cross-sectional area $A_k$ (Fig. \ref{Fig5}g) develops into a sequence of fictitious Fermi pockets that increment by discrete areal values $A_q$ for extremal cyclotron orbits that propagate discretely along $k_\zeta$ (Fig. \ref{Fig5}h). This is distinct from Bragg scattering via the \moire \qvector (\textit{e.g.}, via magnetic breakdown) and instead is similar to a coherent, local hopping process in a dimension orthogonal to physical space. Ultimately, this provides a natural hypothesis for explaining the complex Fermiology in STS(3,90\degree), as generated by the ordered extension of the simple 3-band Fermi surface of monolayer \TaS into a synthetic dimension that generates dense combs of de Haas-van Alphen oscillation frequencies with approximate linear spacing (Figs. \ref{Fig5}i,j; see SI for further discussion).

To date, most incommensurate metals have exhibited low mobilities, likely due to disorder from more complex formation energetics compared to crystalline materials \cite{Bancel1985}. Rare exceptions with high mobilities are aptly described by material-specific models that approximate weak incommensurability effects in terms of Bragg scattering and magnetic breakdown \cite{Razavi1979, Batalla1982, Everson1987, Hill1997, Kawamoto2003, Kawamoto2006a, Kawamoto2006b}. Such models cannot describe the present bulk \moire metals due to their high mobilities (with mean-free paths orders of magnitude larger than $\lambda_m$; see SI) and their pronounced \moire modulations, thus necessitating our generalizable superspace theoretical framework. In addition, because of the homogeneity of their \moire \textit{q}-vectors and the sizes of their Fermi pockets, bulk \moire metals act as ideal material platforms for higher-dimensional Fermiology in comparison to 2D \moire semimetals and semiconductors (see SI for discussion). In total, we propose that the Fermi surfaces of bulk \moire metals can be understood as ($3 + 1$)D Fermi surfaces with $3$ continuous spatial dimensions ($k_x$, $k_y$, $k_z$) and $1$ discrete synthetic dimension ($k_\zeta$).

\begin{flushleft}\textbf{Discussion}\end{flushleft}

Dimensionality is an idea that is pervasive throughout physics and is often associated with the concept of holography, which captures the relationship between a complete description of a higher-dimensional system that emerges from a lower-dimensional encoded representation. Holography is foundational to many fields of physics, including its use in classical optics to produce projected 3D images from 2D films \cite{Schnars1994}, in holographic dualities to describe the correspondence between gravitational and quantum field theories \cite{Susskind1995, Bousso2002}, and in quantum error correction to describe the emergence of fault-tolerant logical qubits from their physical qubits encodings \cite{Preskill, Nielsen2010, Pastawski2015}. Similarly, we propose that holography describes the relationship between certain quantum phenomena of higher-dimensional superspace crystals (\textit{i.e.}, emergent space) encoded in the lower-dimensional lattice degrees of freedom of tunable aperiodic composite crystals (\textit{i.e.}, physical space; see SI for discussion). The present work demonstrates how different synthesis conditions can effectively program the aperiodic lattice of bulk \moire materials in ways that encode different superspace crystal lattices. High-resolution quantum oscillation studies evidence the emergent higher-dimensional Fermiology of these superspace crystals and provide a quantitative fingerprint for the incommensurate lattice structure itself.

Looking forward, we envision many directions for realizing physical phenomena that harness holographic relationships (see SI for discussion) \cite{Ozawa2019, Boada2012, Celi2014, Lustig2019}. A rich landscape of theoretical work has predicted higher-dimensional topological phases with exotic bulk-boundary correspondences \cite{Sugawa2018, Lohse2018, Zilberberg2018}, as well as unconventional superconductivity that breaks higher-dimensional symmetries \cite{Sakai2017, Cao2020}, phenomena traditionally confined to abstract models in four or more spatial dimensions. Broadly, this work demonstrates a powerful new route towards realizing higher-dimensional quantum phases of matter by targeting aperiodic composite materials capable of encoding the requisite emergent superspace lattices for bridging abstract theoretical predictions with experimentally accessible material platforms.

\clearpage

\noindent \textbf{Acknowledgments:} We thank S. Y. Frank Zhao, Paul Neves, Alex Mayo, Eun Sang Choi, Peter M{\"u}ller, Julius Oppenheim, Nilanjan Chatterjee, Herv{\'e} Rezeau, Jordan Cox, and Charlie Settens for their technical support of this project and for useful discussions. We thank the MIT Chemistry Department X-Ray Diffraction Facility, the MIT Electron Microprobe Facility, and the National High Magnetic Field Laboratory for making this work possible.  This work was funded, in part, by the Gordon and Betty Moore Foundation EPiQS Initiative, Grant No. GBMF9070 (synthesis instrumentation and computation), the US Department of Energy (DOE) Office of Science, Basic Energy Sciences, under award DE-SC0022028 (material development), the Office of Naval Research (ONR) under award N000142412407 (material analysis), and ARO grant no. W911NF-16-1-0034 (measurement technique development) to J.G.C. K.P.N. acknowledges support from the MIT Pappalardo Fellowship in Physics. Support for the theory–experimental collaboration reported here was provided by the Air Force Office of Scientific Research (AFOSR), under award FA9550-22-1-0432 to L.F. and J.G.C. F.G. is grateful for the financial support from the Swiss National Science Foundation (Postdoc.Mobility Grant No. 222230). A portion of this work was performed at the National High Magnetic Field Laboratory, which is supported by the National Science Foundation Cooperative Agreement No. DMR-2128556 and the State of Florida. This work was additionally carried out in part through the use of MIT.nano's facilities. The imaging work was supported by the STC Center for Integrated Quantum Materials, NSF Grant No. DMR-1231319, and it was performed in part at the Harvard University Center for Nanoscale Systems (CNS); a member of the National Nanotechnology Coordinated Infrastructure Network (NNCI), which is supported by the National Science Foundation under NSF award no. ECCS-2025158.

\vspace{5 mm}

\noindent \textbf{Author Contributions:} K.P.N. and J.G.C. designed the experiments. K.P.N. discovered the compounds used for the study with input on their synthesis from T.S. and J.G.C. K.P.N. performed structural characterization of the compounds and fabricated torque magnetometry and magnetotransport samples used in measurements. K.P.N. performed torque magnetometry and magnetotransport measurements with measurement support from A.C. and J.P.W. and with technical support from D.G. High-angle annular dark-field scanning transmission electron microscopy measurements were performed by A.A. with support from J.G., A.J.A., and D.C.B. Theoretical modeling and calculations were performed by N.P. and F.G. with guidance from L.F. and with input from K.P.N., A.C., and J.G.C. All authors discussed the results and contributed to the writing of the manuscript.

\pagebreak

\noindent \textbf{Methods}

\noindent \textbf{Single-crystal synthesis} \\
For full details of each growth recipe, see SI. Briefly, SrS powder, Ta powder, S pieces, and either SrCl$_2$ beads or SrBr$_2$ powder were combined in an Ar glovebox, sealed at approximately $5 \times 10^{-7}$ Torr without air exposure, and grown under different temperature sequences for each \moire superlattice, as detailed fully in the SI. In total, the \moire superlattices in \STSnphi compounds are controlled by the growth's precursor stoichiometry (SrS:Ta:S), its temperature sequence, and its salt catalyst (SrCl$_2$ or SrBr$_2$).

\vspace{5 mm}

\noindent \textbf{X-ray diffraction} \\
Wide-angle X-ray scattering measurements were performed using the facilities of MIT.nano. Thin ($< 100 \: \mu \text{m}$) plate-like crystals were affixed to etched SiN$_x$ windows using silicone vacuum grease. Samples were analyzed using a SAXSLAB Retro-F instrument at room temperature under x-ray irradiation directed along the crystal's out-of-plane \caxis. Single-crystal X-ray diffraction measurements were performed in the MIT Chemistry Department X-Ray Diffraction Facility. Crystals were mounted on a MiTeGen loop using paratone oil. Samples were cooled to 100 K using liquid nitrogen, and analyzed using a Bruker Single-Crystal X-Ray Diffractometer.

\vspace{5 mm}

\noindent \textbf{Wavelength-dispersive X-ray spectroscopy} \\
Wavelength-dispersive X-ray spectroscopy measurements were performed in the MIT Electron Microprobe Facility. Flat plate-like single crystal samples were chosen of each compound and affixed to a glass slide using carbon tape. This glass slide was then coated with a thin conductive film in order to eliminate charging artifacts, and the samples were analyzed using a JEOL JXAS-8200 Superprobe capable of performing scanning electron microscopy imaging and wavelength-dispersive spectroscopy measurements of the the stoichiometries of crystals. Prior to each session of stoichiometric measurements performed on all of the single crystals affixed to a single glass slide, the tool was calibrated against a set of standard samples with known stoichiometric ratios of Sr, Ta, S, O, and Cl.

\vspace{5 mm}

\noindent \textbf{Torque magnetometry} \\
Torque magnetometry measurements were performed at Cell 9 (main text) and Cell 6 (SI) of the National High Magnetic Field Laboratory and in a commercial 14 T in-house cryostat (SI). Torque was measured using piezoresistive cantilevers (SCL SensorTech PRSA-L300) with an excitation amplitude of 25 mV applied to the top and bottom nodes of a resistive bridge including the sample piezoresistive cantilever and a room-temperature variable resistor used for balancing the bridge. de Haas-van Alphen oscillations of the torque magnetization were analyzed after a smooth polynomial background (often quadratic or quartic) was removed from the raw torque signal, and the background-subtracted signal was Fourier transformed with the use of a Blackman windowing function used to suppress spectral leakage artifacts and higher harmonics. Details of this analysis and the role of the field window, the windowing function, and the background subtraction in the analyses are discussed in the SI.

\vspace{5 mm}

\noindent \textbf{Magnetotransport measurements} \\
Longitudinal transport measurements were performed at Cell 6 of the National High Magnetic Field Laboratory. Current excitation of 1-2 mA was applied along the crystal's \aaxis, and voltage modulations were detected using the standard a.c. technique with a lock-in amplifier.

\vspace{5 mm}

\noindent \textbf{Theoretical calculations and modeling} \\
For full details about the momentum-space network model derived from a Lifshitz-Kosevich treatment of quasi-2D metals with incommensurate 1D \moire modulations, see SI.

\vspace{5 mm}

\noindent \textbf{Data Availability} \\
The data that support the findings of this study are available from the corresponding author on reasonable request.

\vspace{5 mm}

\noindent \textbf{Competing Interests} \\
The authors declare no competing interests.

\bibliography{bibliography}

\pagebreak

\pagebreak

\pagebreak

\pagebreak

\pagebreak

\pagebreak

\end{document}